\documentclass[english,APS,PRL,twocolumn]{revtex4-1}
\usepackage[T1]{fontenc}
\usepackage[latin9]{inputenc}
\usepackage{graphicx}
\usepackage{esint}

\makeatletter

\@ifundefined{textcolor}{}
{%
 \definecolor{BLACK}{gray}{0}
 \definecolor{WHITE}{gray}{1}
 \definecolor{RED}{rgb}{1,0,0}
 \definecolor{GREEN}{rgb}{0,1,0}
 \definecolor{BLUE}{rgb}{0,0,1}
 \definecolor{CYAN}{cmyk}{1,0,0,0}
 \definecolor{MAGENTA}{cmyk}{0,1,0,0}
 \definecolor{YELLOW}{cmyk}{0,0,1,0}
}

\makeatother

\usepackage{babel}
\begin{document}

\title{Position-Dependent Local Detection Efficiency in a Nanowire Superconducting
Single-Photon Detector }

\author{J.J. Renema}

\email{renema@physics.leidenuniv.nl}

\author{Q. Wang}

\affiliation{Leiden Institute of Physics, Leiden University, Niels Bohrweg 2,
2333 CA Leiden, the Netherlands}

\author{R. Gaudio}

\affiliation{COBRA Research Institute, Eindhoven University of Technology, P.O.
Box 513, 5600 MB Eindhoven, The Netherlands}

\author{I. Komen}

\affiliation{Leiden Institute of Physics, Leiden University, Niels Bohrweg 2,
2333 CA Leiden, the Netherlands}

\author{K. op 't Hoog}

\affiliation{COBRA Research Institute, Eindhoven University of Technology, P.O.
Box 513, 5600 MB Eindhoven, The Netherlands}

\author{D. Sahin}

\affiliation{COBRA Research Institute, Eindhoven University of Technology, P.O.
Box 513, 5600 MB Eindhoven, The Netherlands}

\author{A. Schilling}

\affiliation{Physics Institute of the University of Zurich, Winterthurerstr. 190,
8057 Zurich, Switzerland}

\author{M.P. van Exter}

\affiliation{Leiden Institute of Physics, Leiden University, Niels Bohrweg 2,
2333 CA Leiden, the Netherlands}

\author{A. Fiore}

\affiliation{COBRA Research Institute, Eindhoven University of Technology, P.O.
Box 513, 5600 MB Eindhoven, The Netherlands}

\author{A. Engel}

\affiliation{Physics Institute of the University of Zurich, Winterthurerstr. 190,
8057 Zurich, Switzerland}

\author{M.J.A. de Dood}

\affiliation{Leiden Institute of Physics, Leiden University, Niels Bohrweg 2,
2333 CA Leiden, the Netherlands}
\begin{abstract}
We probe the local detection efficiency in a nanowire superconducting
single-photon detector along the cross-section of the wire with a
spatial resolution of 10 nm. We experimentally find a strong variation
in the local detection efficiency of the device. We demonstrate that
this effect explains previously observed variations in NbN detector
efficiency as function of device geometry.
\end{abstract}
\maketitle
Nanowire superconducting single-photon detectors (SSPDs) consist of
a superconducting wire of nanoscale cross-section \cite{Goltsman2001},
typically 4 nm by 100 nm. Photon detection occurs when a single quantum
of light is absorbed and triggers a transition from the superconducting
to the normal state. SSPDs have high efficiency, low jitter, low dark
count rate and fast reset time \cite{Natarajanrevie}, and are therefore
a key technology for, among others, quantum key distribution \cite{Takesue2007},
interplanetary communication \cite{Boroson2009} and cancer research
\cite{Gemmell2013}.

Although progress has been made recently, the underlying physical
mechanism responsible for photon detection on the nanoscale is still
under active investigation. A combination of theory \cite{Bulaevskii2011,Bulaevskii2012},
experiments \cite{RenemaPRB,RenemaPRL,Luschewidthdep} and simulations
\cite{Engelpreprint,EngelarXiv2} on NbN SSPDs indicates that the
absorption of a photon destroys Cooper pairs in the superconductor
and creates a localized cloud of quasiparticles that diverts current
across the wire. This makes the wire susceptible to the entry of a
superconducting vortex from the edge of the wire. Energy dissipation
by this moving vortex drives the system to the normal state.

An important and unexpected implication of this detection model is
a nanoscale position variation in the photodetection properties of
the device. The conditions at the entry point of the vortex determine
the energy required for it to cross the wire. This causes photons
absorbed close to the edge to have a local detection efficiency (LDE)
\footnote{We define the LDE as the conditional probabilty of a detection event,
given photoabsorption at a particular location%
} compared to photons absorbed in the center of the wire \cite{EngelarXiv2}.
This effect has practical implications for the operation of SSPDs,
since it represents a potential limitation of the detection efficiency.
In addition, SSPDs have been proposed for nanoscale sensing, either
in a near-field optical microscope configuration \cite{QiangOE} or
as a subwavelength multiphoton probe \cite{Bitauld2010}, where this
effect would be of major importance for the properties of such a microscope.
While this effect has been predicted theoretically, clear experimental
evidence is missing.

In this work, we experimentally explore the nanoscale variations in
the intrinsic response of the detector. We spatially resolve the LDE
with a resolution of approximately 10 nm, better than $\lambda/50$,
using far-field illumination only. We find that our results are qualitatively
consistent with numerical simulations \cite{EngelarXiv2,Engelpreprint}.
Our results provide excellent quantitative agreement with experiments
that indicate a polarization dependence in the LDE that was hitherto
not understood \cite{Anant2008}.

The key technique used in this work is a differential polarization
measurement that probes the IDE of the detector (see Figure 1). The
technique is based on the fact that polarized light is absorbed preferentially
in different positions for the two orthogonal polarizations, due to
differences in boundary conditions. Using this technique, we achieve
selective illumination of either the sides or the middle of the wire,
which we use to probe the intrinsic photodetection properties of our
device on the nanoscale.

However, it is well known that changing the polarization results in
a change in overall optical absorption probability \cite{Anant2008,Driessenpol,Driessenabsorber,Vermapoldep,Yamashita2013}.
In order to correct for this effect, we must separate the probability
that a photon is absorbed from the probability that an absorbed photon
causes a detection event. To separate changes in optical absorption
from the intrinsic LDE effects which are presently of interest, we
use quantum detector tomography (QDT) \cite{RenemaPRL,Lundeen2008,Feito2009,Akhlaghi2009,Akhlaghi2009a,Lundeen,Brida2011,RenemaPRA,Renema2012,RenemaPRB,ZwillerOE}. 

QDT records the detector response to a set of known quantum states
of light and distills from these measurements the detection probability
for different photon numbers. As we showed previously \cite{Renema2012},
this procedure allows us to unambiguously separate the intrinisic
one-photon detection probability $p_{1}$ from the probability $\eta$
that a photon is absorbed %
\footnote{See Supplemental material for details.%
}. We found that $\eta$ is almost independent of detector bias current
and that the value is consistent with the geometric area of the detector.
Hence, we identify $p_{1}$ with the detection probability conditional
on photon absorption, which we dub the internal detection efficiency
(IDE). By construction, $\mathrm{IDE}=\int\mathrm{LDE}(x)A(x)dx/\int A(x)dx,$
where the integral runs along the cross-section of the wire and $A(x)$
represents the absorption probability density. 

\begin{figure}[h]
\begin{centering}
\includegraphics[width=8.8cm]{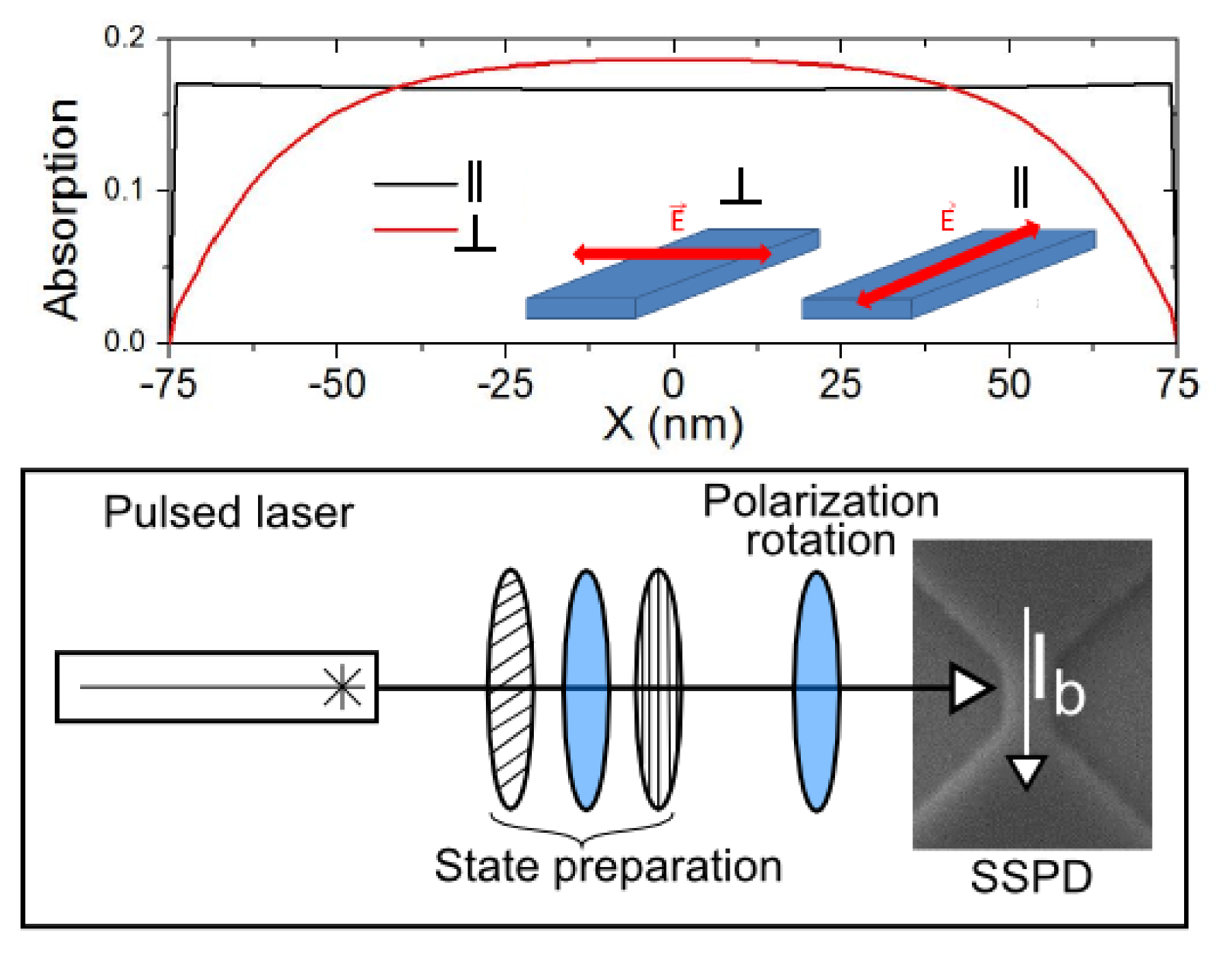}
\par\end{centering}

\protect\caption{Sketch of our experiment. \emph{Top: }Absorption as a function of
position in the wire for parallel ($\parallel$) and perpendicular
($\perp$) polarizations, calculated with an FDTD method (see text).
\emph{Inset}: Sketch showing the two polarizations. The red arrow
represents the polarization of the electric field. \emph{Bottom}:
Experimental setup. Our laser pulses are tuned in intensity by a variable
attenuator consisting of two crossed polarizers and a $\lambda/2$
wave plate. Polarization is set by an additional $\lambda/2$ wave
plate. The image is a SEM micrograph of a detector nominally identical
to the one used in this experiment.}
\end{figure}

\begin{figure}
\begin{centering}
\includegraphics[width=8.8cm]{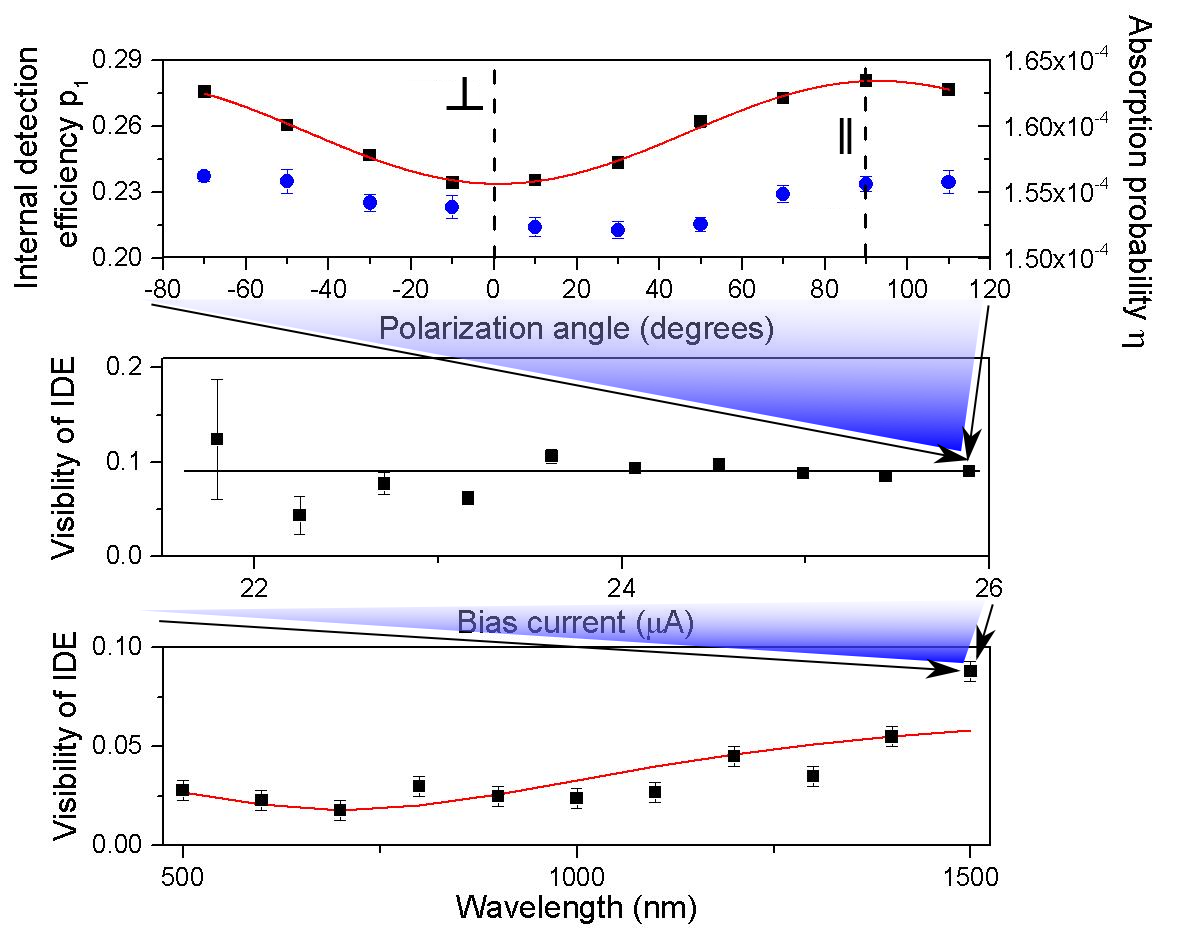}
\par\end{centering}

\protect\caption{Experimental results on the polarization dependence of the internal
detection efficiency (IDE). Each graph represents one data point in
the graph below it. \emph{Top panel, black squares: }IDE as a function
of polarization at an excitation wavelength of 1500 nm at $I_{b}=25.8\ \mathrm{\mu A}$.
The red line is a sine fit. \emph{Top panel, blue circles: }Absorption
probability as a function of polarization. \emph{Middle panel: }IDE
as a function of bias current. We find that the IDE is independent
of bias current. The black line represents the weighted average of
our measurements at different bias currents. \emph{Bottom panel:}
IDE as a function of illumination wavelength. We show the fit of the
observed polarization-dependent IDE to the position-dependent local
detection efficiency (LDE).}
\end{figure}

We perform our experiments on a 100 nm long, 150 nm wide NbN bridge
patterned from a 5 nm-thick NbN film sputtered on a GaAs substrate
\cite{Gaggero2010} ($I_{c}=28\ \mathrm{\mu A})$. We apply a bias
current and read out the detector with a bias-tee to separate high-frequency
detection pulses from the DC bias current. The resulting pulses are
fed to a series of RF amplifiers and a pulse counter. At each combination
of bias current, photon energy and polarization, we record the detector
count rate as a function of input intensity. The probe states were
prepared by a broadband pulsed laser (Fianium, repetition rate 20
MHz) out of which we select a narrow wavelength band with dielectric
filters \cite{RenemaPRL}. We prepare the desired intensity and polarization
by first attenuating the light with a combination of two crossed polarizers
and a half-wave plate, and then setting the polarization with an additional
wave plate (see Figure 1) %
\footnote{While in principle it is possible to achieve the desired combination
of polarizations and intensities using two independently rotating
polarizers, we found that the effects of wedge in the polarizers preclude
this solution. %
}.

Figure 2 shows our measured IDE and $\eta$, as function of polarization
and wavelength. The top panel shows the IDE of our device as a function
of the polarization of light with $\lambda=1500\ \mathrm{nm}$ wavelength.
Our experiments show that the IDE and absorption probability oscillate
in phase when the polarization is rotated, with a minimum at perpendicular
polarization and a maximum at parallel polarization. This demonstrates
that absorption of TM-polarized photons is less likely to result in
a detection. This polarization is absorbed preferentially in the middle
of the wire. Our result therefore confirms earlier, preliminary results
\cite{Anant2008,Maingault,sahin2013waveguide} that suggested that
the edges of the detector are more efficiently photodetecting than
the center of the wire. The error bars in this panel correspond to
the standard deviation of a series of independent experiments. The
middle panel of Figure 2 shows that the visibility of the IDE $\mathrm{V_{IDE}}=(p_{max}-p_{min})/(p_{max}+p_{min})$
is independent of bias current. We can therefore associate one visibility
to a particular illumination wavelength. For $\lambda=1500$ nm we
find $\mathrm{V_{IDE}}=0.09$. The bottom panel shows the wavelength
dependence of the visibility. We find that longer wavelengths have
higher visibility in the IDE. In this panel, we show a fit to our
position-dependent LDE, which we will discuss below.

In the second part of this work, we will discuss our reconstruction
of the LDE profile from these measurements. We make use of the fact
that the IDE is given by the LDE multiplied by the optical absorption
probability, integrated across the wire. Our strategy is to take the
absorption profiles as given - since they are well studied - and to
take the LDE profile as a free parameter and fit it to our experimental
data. 

To calculate the optical absorption distributions, we perform a series
of numerical simulations at different wavelengths using a finite-difference
time domain (FDTD) method (RSOFT Fullwave). We perform a 2D simulation,
and compute the electric field in a 150 nm wide, 5 nm thick NbN wire
on a semi-infinite GaAs substrate and an 80 nm thick HSQ layer on
top of the NbN wire. The refractive index of NbN deposited film on
GaAs is derived from spectroscopic ellipsometry measurements \cite{donduthesis}
\footnote{In these calculations, we neglect the effect of the tapered parts
of the bridge because they have little influence on the absorption
in the central, photodetecting section. See Supplemental material
for details.%
}. In Figure 1, the result of this calculation is shown for $\lambda$
= 1500 nm. 

In order to combine information from different wavelengths - which
is required to find the LDE - we must posit some relation between
photon energy and detection probability %
\footnote{This calculation is described in detail in the Suplemental Material%
}. Our ansatz is motivated by the experimental observation that for
low detection probabilities, the detection probability depends exponentially
on bias current, by our earlier work on the energy-current relation
in SSPDs, and on the numerical simulations described below \cite{RenemaPRL,RenemaPRB,EngelarXiv2,Engelpreprint}.
We model the bias current dependence of the LDE as $\mathrm{LDE}(x)=\min\{1,\exp(I_{b}-I_{th}(x))/I^{\star}\}$,
with $I_{th}(x)=I_{c}-\gamma'(x)E,$ where $I_{b}$ is the applied
bias current, $I_{c}$ is the critical current, $E$ is the photon
energy, and $I^{\star}=0.65\ \mathrm{\mu A}$ is an experimentally
determined current scale. $\gamma'(x)$ is the local energy-current
interchange ratio, which parametrizes the detection probability of
the wire at different excitation wavelengths. These simple, empirical
expressions enable us to combine our wavelength- and, polarization-dependent
IDE measurements, and compute the LDE.

\begin{figure}
\includegraphics[width=8.8cm]{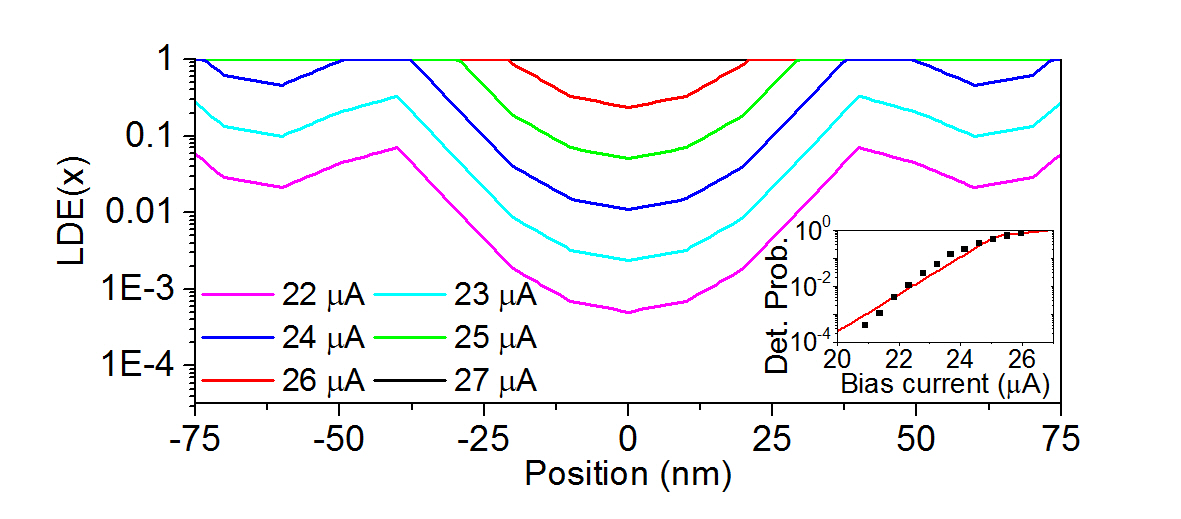}

\includegraphics[width=8.8cm]{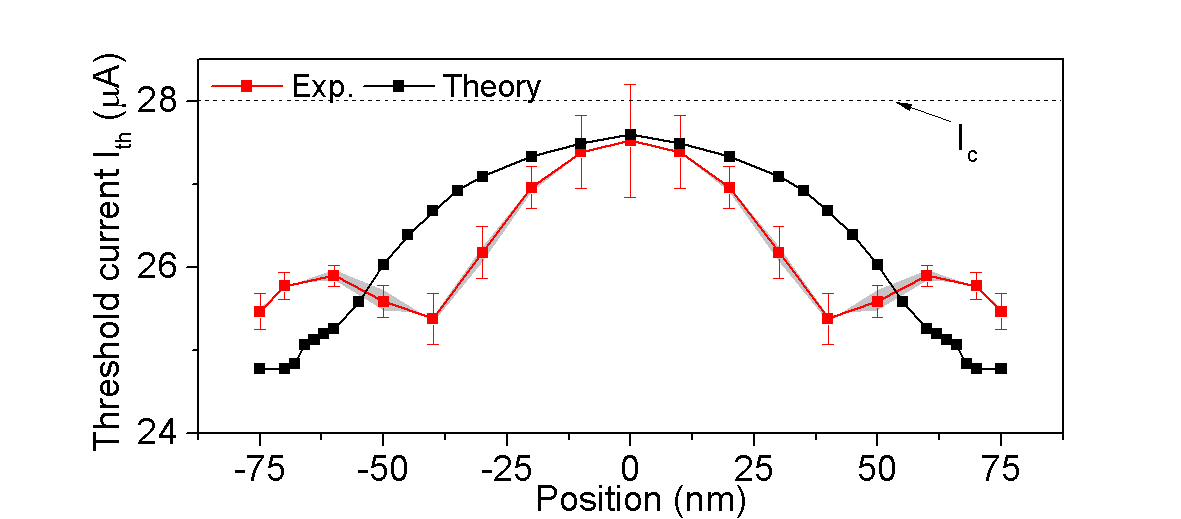}

\protect\caption{Local detection efficiency (LDE) and threshold current in an SSPD,
for $\lambda=1500\ \mathrm{nm}$. \emph{Top panel}: observed positional
variations in the LDE of an SSPD, as a function of bias current. \emph{Inset}:
IDE, as a function of bias current. \emph{Bottom panel: }Threshold
current $I_{th}$. The red curve shows the experimentally observed
value, the black curve shows the theoretical value obtained from numerical
modelling of the detection event. }
\end{figure}

The top panel of Figure 3 shows the resulting LDE profile for different
bias currents, for a wavelength of 1500 nm. We obtained this result
by fitting $\gamma'(x)$ to the experimentally determined visibility
of the polarization-dependent IDE in the range $\lambda=500-1500$
nm, using the calculated absorption profiles. We find that the LDE
has a high value at the edges of the wire, up to a point roughly 40
nm from the edge. From there, the detection efficiency decreases;
it is reduced by two orders of magnitude at the center of the wire.
The LDE is current-dependent, but saturation sets in around $I_{b}=25.5\ \mu\mathrm{A}$
($I_{b}/I_{c}=0.91)$. 

The inset in the top panel of Figure 3 shows the effect of this saturation
on the IDE. There are three regimes: a rolloff regime, where the detection
property depends exponentially on bias current, a plateau regime at
high currents, where LDE = 1, implying IDE = 1, and an intermediate
regime of slowly increasing detection probability. These regimes are
marked by dashed lines in the inset. In the middle regime, parts of
the detector are fully photodetecting, while other parts are still
in a fluctuation-assisted regime \cite{VodolazovPRB}. The variations
between the experimental data and the values calculated from our fit
are less than a factor of 2, which demonstrates the self-consistency
of our results. We have also confirmed that the observed LDE profile
reproduces the fact (see Figure 2, middle panel) that in our measurement
range, the visibility of the polarization-dependent IDE does not depend
on the applied bias current. From these observations, we conclude
that our description is able to explain all of the available experimental
data regarding polarization and bias current dependence of the photoresponse
of this detector.

We estimate the resolution in our measurement by varying the distance
$\Delta x$ at which we specify $\gamma'(x).$ In particular, we find
that for $\Delta x\approx w/2$, we are unable to explain our experimentally
observed data. Therefore, we conclude that the LDE profile varies
with a length scale smaller than the dimensions of the wire. We find
that we achieve the best fit at $\Delta x$ = 10 nm. From this, we
conclude that this is the resolution of our experiment.

The bottom panel of Figure 3 shows the threshold current $I_{th}(x)$.
The red curve shows the experimental value, from which the $\mathrm{LDE(x)}$
shown in the top panel is derived. The black curve in the bottom panel
of Figure 3 shows an independent, ab initio calculation of the threshold
current based on the model described below. This ab initio calculation
of the position-dependent detection probability is based on a numerical
model \cite{EngelarXiv2,Engelpreprint} that determines the threshold
current for the detection of an absorbed photon of a given wavelength
using a combination of quasiparticle diffusion, current displacement
and vortex entry %
\footnote{See Supplemental material for details.%
}. We find reasonable agreement between our observed experimental results
and the theoretical values. From this, we conclude that this model
captures the essential physics of the detection process in SSPDs. 

Our ab initio calculation gives a physical explanation for the enhanced
efficiency at the edges of the wire in terms of our microscopic model.
Comparing a photon absorption in the center of the wire to one at
the edge, there are two differences. First, for an absorption event
at the edge, the current density at the edge of the wire is reduced,
due to the reduction in the number of superconducting electrons $n_{s}$.
However, this is more than compensated by the reduction of the vortex
self-energy, which is proportional to $n_{s}$. Vortices enter more
easily when the superconductivity is weakened at their entry point,
and that makes the detector more efficient at the edges.

We note that there is some disagreement in theoretical literature
about the predicted shape of the LDE curve. The alternative model
of Zotova \emph{et al. }\cite{ZotovaARXIV}\emph{, }which is based
on the Ginzburg-Landau formalism, naturally takes into account vortex
entry. However, it disregards quasiparticle diffusion and implements
a hotspot with hard boundaries. The results from this model disagree
strongly with our experimental results: there, a W-shaped threshold
current profile is predicted, with threshold currents at the edges
almost as high as in the center of the wire. The discrepancy between
the models occurs precisely at the point where their 'hard' hotspot
touches the edge of the wire. We speculate that both models, if refined
more, will likely converge.

\begin{figure}
\includegraphics[width=8.8cm]{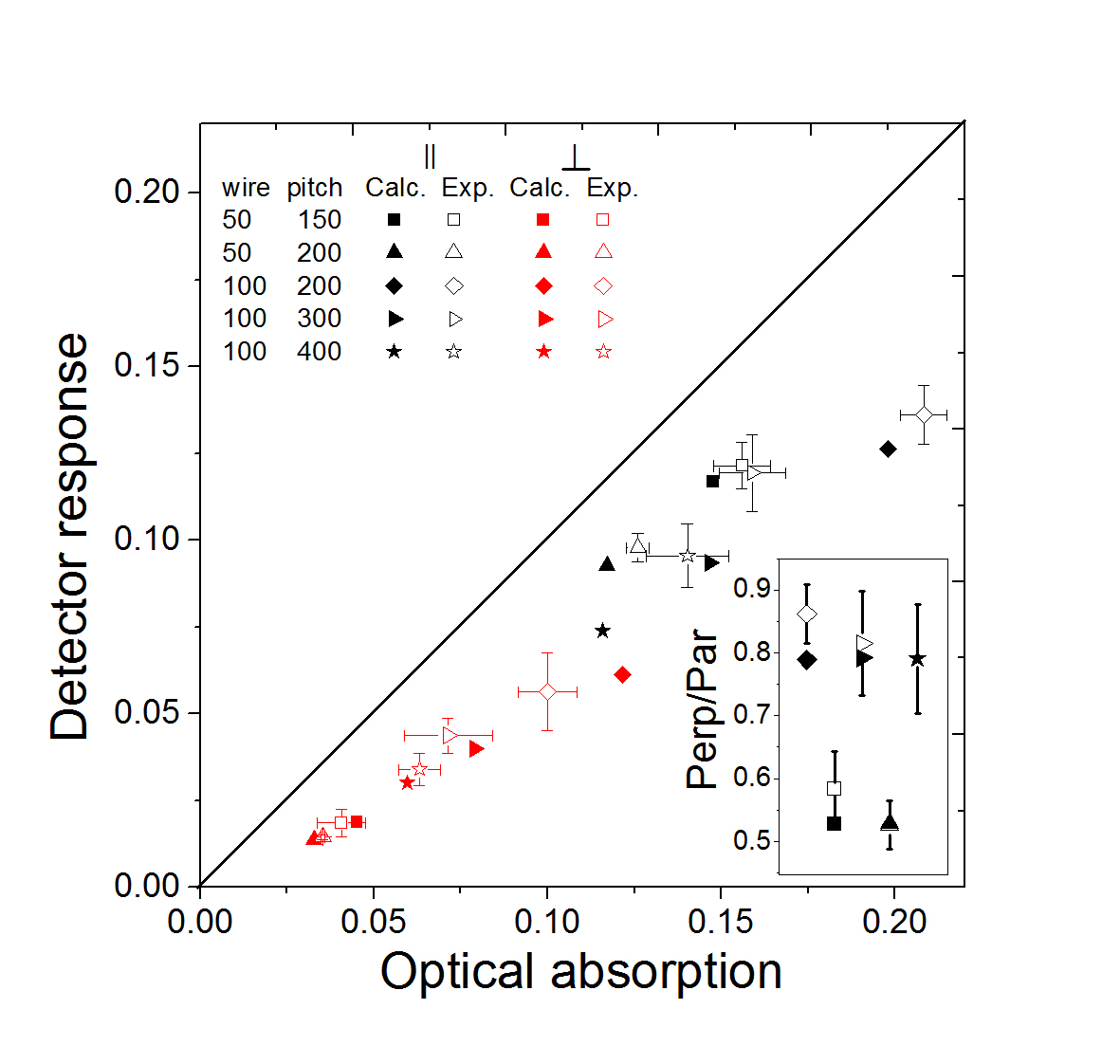}

\protect\caption{Internal detection efficiency (IDE) of a set of SSPD meander devices.
The solid symbols represent our calculation, and the open symbols
represent experimental data from \cite{Anant2008}. Red symbols represent
perpendicular ($\perp$) polarization, and black symbols represent
parallel ($\parallel$) polarization. The shape of the symbol represents
the wire and pitch of the detector (see legend). The error bars on
the experimental data represent the spread in properties between detectors
of the same design. The diagonal line represents the case IDE = 1.
\emph{Inset} Ratio of overall absorption for the two polarizations. }
\end{figure}

In the final part of this work, we discuss the implication of these
results to the construction and operation of SSPD devices. In the
work of Anant \emph{et al.} \cite{Anant2008}, the IDE of a series
of meander SSPDs of different wire width and pitch (wire separation)
was measured. Using similar methods as described above, we compute
the position-dependent optical absorption of such structures for the
film properties given in that work. We use the observed LDE curve
from our experiment, and compute the IDE and overall detection probability
of such devices, using the same expressions which we used for our
sample to obtain the LDE from the IDE and the optical absorption %
\footnote{We account for the differences in width by considering the edges of
our IDE profile. See Supplemental material for details%
}.

Figure 4 shows that this prediction agrees with the data. The calculation
confirms the general claim made by Anant \emph{et al.} that parallel
polarization has a higher IDE than perpendicular. Beyond that, we
are also able to compute the IDE for each device independently. The
inset of Figure 4 shows the ratio of the overall efficiency for the
two polarizations, which factors out the optical absorption as well
as - to first order - the effect of bias current. It is therefore
the most direct test of our IDE. We find excellent agreement. This
demonstrates that we have achieved quantitative understanding of the
internal properties of an SSPD. It also demonstrates that our results
are neither limited to the SSPD on which we performed our experiment
nor by our single-wire geometry. 

To obtain a nonunity IDE, we must assume that the highly efficient
detectors reported in \cite{Anant2008} were not biased to their critical
current. It is well known that the presence of current crowding in
the bends of a wire can cause reduction of the device critical current
by as much as 40\% \cite{Henrichcurrentcrowding}, with typical values
of 10-20\% \cite{Kerman2006}. We have assumed $I_{b}/I_{c}\approx0.9$
for all devices to produce Figure 4. This demonstrates that our results
are relevant for the kind of SSPDs which are used for applications,
at the typical currents at which they are operated. It also shows
that our results can be used as an all-optical method for measuring
the amount of current crowding in these devices. Since the differences
between the two optical absorption profiles become larger at longer
wavelengths, we expect our results to be particularly relevant for
the engineering of SSPDs at mid-infrared wavelengths.

In conclusion, we have demonstrated that the local detection efficiency
of an SSPD depends on the position along the cross-section at which
the photon is absorbed. We have probed this effect with a resolution
of approximately 10 nm, and found agreement with theoretical calculations
done in the context of the diffusion-based vortex crossing model.
From this, we conclude that this model contains the essential features
for a complete microscopic picture of the detection model in SSPDs.
We have compared these predictions of our work to results reported
on devices used for applications and found good agreement, demonstrating
the relevance of our results for SSPD engineering. These results enable
quantitative modeling of the internal properties of SSPDs.
\begin{acknowledgments}
We thank D. Boltje for technical assistance, F. Schenkel for assistance
with the experimental apparatus, M. Frick for assistance with programming
and E.F.C. Driessen for critical reading of the manuscript. This work
is part of the research programme of the Foundation for Fundamental
Research on Matter (FOM), which is financially supported by the Netherlands
Organisation for Scientific Research (NWO). It is also supported by
NanoNextNL, a micro- and nanotechnology program of the Dutch Ministry
of Economic Affairs, Agriculture and Innovation (EL\&I) and 130 partners,
and by the Swiss National Science Foundation grant no. 200021\_146887/1.
DS was supported by the Marie Curie Actions within the Seventh Framework
Programme for Research of the European Commission, under the Initial
Training Network PICQUE, Grant No. 608062.
\end{acknowledgments}

\end{document}